\documentclass[twocolumn,showpacs,aps]{revtex4}
\usepackage{graphicx}
\usepackage{bm}% bold math
\usepackage{amsmath}
\usepackage{amsfonts}
\usepackage{amssymb}
\usepackage[latin1]{inputenc}
\usepackage{color}

\newcommand{\ket}[1]{|#1\rangle}

\newcommand{\scalar}[2]{\langle#1|#2\rangle}

\def\Tr{{\rm Tr}}
\def\sys{{\rm sys}}
\def\Env{{\rm Env}}

\begin{document}

\title{Decoherence in one-dimensional Quantum Walk}

\author{Mostafa Annabestani}
\email{Annabestani@modares.ac.ir}\affiliation{Dept. of Physics, Basic Sciences Faculty, Tarbiat Modarres University, Tehran, Iran}
\author{Seyed Javad Akhtarshenas}
\email{akhtarshenas@phys.ui.ac.ir} \affiliation{Dept. of
Physics, University of Isfahan, Isfahan, Iran\\
Quantum Optics Group, University of Isfahan, Isfahan, Iran}
\author{Mohamad Reza Abolhassani}
\altaffiliation{Dept. of Physics, Basic Sciences Faculty, Tarbiat Modarres University, Tehran, Iran}

\date{\today}
\begin{abstract}
In this paper we study decoherence in the quantum walk on the
line. We generalize the method of decoherent coin quantum walk,
introduced by Brun et al [Phys. Rev. A {\bf 67}, 32304 (2003)].
Our analytical expressions are applicable for all kinds of
decoherence. As an example of the coin-position decoherence, we
study the broken line quantum walk and compare our results with
the  numerical one. We also show that our analytical results
reduce to the  Brun formalism when only the coin is subjected to
decoherence.
\end{abstract}

% PACS, the Physics and Astronomy  Classification Scheme.
%03.67.-a    Quantum information
%03.67.Mn    Quantum entanglement production, characterization, and manipulation
%03.65.Ud    Entanglement and quantum nonlocality (e.g. EPR paradox, Bell's inequalities, GHZ states, etc.)

 %Use showkeys class option if keyword display desired

\maketitle

\section{Introduction}
The quantum walk (QW) is the quantum analogue of the classical
random walk (CRW). Notable differences between the QW and the CRW
are the quadratic dependency of variance on the number of steps
and the complex oscillatory probability distribution in the QW
instead of the linear variance dependency and the binomial
probability distribution in the CRW. These differences between the
QW and the CRW have been used to present several quantum
algorithms in order to solve some specific problems
\cite{Shenvi03,Ambainis03,AKR,Ambainis05,Childs04,Tulsi08} with
performances better than the best known classical versions.
Recently Childs has shown  that the universal computation can be
performed by QW \cite{Childs2009} and it is, therefore, another
witness of QW importance. Two types of QW have been introduced as
the quantum mechanical extension of CRW: \textit{discrete}
\cite{Nayak} and \textit{continuous time} \cite{FG98}. Both have
the same result in our problems but finding the relation between
them  was an open problem for several years. Recently this
relation has been found \cite{Childs08,Strauch}.
Another classification of QW is based on the network over which the walk takes place. The QW on a line is the simplest possible configuration \cite{Nayak,qw-markov,Konno,Venegas}, but other topologies such as cycles \cite{cycle1,cycle2}, two-dimensional lattices \cite{Mackay,Carneiro,Omar,Amanda,Konno2}, or $n$-dimensional hypercubes \cite{Moore,Marquezino} have also been investigated.
%Although the
%one-dimensional QW is the simplest configuration that has
%attracted much attention \cite{Nayak,qw-markov,Konno,Venegas}, but
%other geometries  are also important and  efforts have been also
%made to investigate the QW on the  cycles \cite{cycle1,cycle2},
%two-dimensional lattices \cite{Mackay,Carneiro,Omar,Amanda,Konno2}
%and $n$-dimensional hypercubes \cite{Moore,Marquezino}.

Quantum entanglement is another important property of quantum
mechanics which has recently attracted much attention in view of
its connection with the QW. The effect of  entanglement of the
coin subspace on the QW \cite{Omar,Venegas,Chaobin}, entanglement
between the coin and position subspaces
\cite{Carneiro,Abal06-ent,annabestani} and the QW as the
entanglement generators
\cite{Goyal,Venegas-entanglementgeneration} are examples of these
studies. Beside of all theoretical studies, experimental
implementation and realization of the QW is another interesting
subject for researchers
\cite{Travaglione,Knight-OC,Dur-prop,Du,Ryan,Zhao-prop}.

In the experimental implementation, the environment effects will
be so important, because in practice,  the preparation of pure
quantum states without interaction with the environment is
impossible, and the environment can disturb the quantum states and
fades the quantum properties. Therefore it is very important to
formulate and quantify the influence of decoherence on the QW and
several valuable researches have been done about the decoherent QW
\cite{Kempe03,KT03,Brun03,Brun2,Dur-prop,Amanda,lopez,deco}.  Due
to the complexity of the analytical calculations, in most of the
previous studies the numerical calculations have been used in
investigation of the effects of decoherence on the QW. Although
analytical expressions have been driven for some particular cases
such as the QW with a coin subject to decoherence \cite{Brun03} or
the weak noises case \cite{Kendon2003}, the general analytical
formulas with the wide range of use are not found  yet. The aim of
this paper is to generalize  the approach of Brun et al
\cite{Brun03} and show that such generalization  is applicable for
all kinds of decoherence in the one-dimensional QW.

This work is organized as follows. Section II gives a brief review
on the one-dimensional QW and decoherence. Section III is devoted
to drive analytical expressions for the first and  second moments
in the presence of general decoherence. Our results have then been
used in section IV to analyze the \textit{coin-position
decoherence} for which separation between the coin noise and the
position noise is impossible and, as an example, an analytical
calculation on the broken line noise has been presented. An
analysis of the \textit{coin decoherence} for which only the coin
is subjected to decoherence is also presented in this section and
it is shown that our results reduce to the Brun et al formula for
the coin decoherence. We summarize our results and present our
conclusions in section V.

\section{Background}\label{background}
From the various methods of studying  the effects of decoherence
on the quantum systems, the Kraus representation is one of the
widely accepted method \cite{Nielsen}. Let us define $H_W$ as the
Hilbert space of our system (Walker) and $H_E$ as the Hilbert
space of environment spanned by $\{\ket{e_n}\}_{n=0}^{m}$ where
$m$ is the dimension of the environment's Hilbert space. In
practice, it is not possible to completely isolate the system from
the environment, therefore in order to find the time evolution of
the system we should consider the time evolution of the whole
system (system+environment), and obtain the state of the system by
tracing  out over the environment's degrees of freedom, i.e.
\begin{equation}\label{rho-sys}
    \rho_{\sys}=\Tr_{\Env} \left( U \rho\, U^\dagger \right).
\end{equation}
Here $U$ acts both on  the system and environment Hilbert spaces.
Without loss of generality we assume that the state of the whole
system is $\rho=\rho_0\otimes|env_0\rangle\langle env_0| $. So we
can write Eq. (\ref{rho-sys}) as
\begin{equation}\label{KR-rho-rep-1}
\rho_{\sys}=\sum\limits_{n = 0}^m {\left\langle {e_n }
\right|U\left| {env_0 } \right\rangle \rho _0 \left\langle {env_0
} \right|U^\dag  \left| {e_n } \right\rangle }  = \sum\limits_{n =
0}^m {E_n \rho_0 E_n^\dag  }
\end{equation}
where $E_n = \left\langle {e_n } \right|U\left| {env_0 }
\right\rangle $, $n=0,1,\cdots,m$, are the so called Kraus
operators.
These operators preserve the trace condition, i.e. $\sum\limits_{n = 0}^m {E_n^\dag E_n}=I$.\\

\subsection{Decoherent one-dimensional QW}
QW is the quantum version of the CRW where instead of the coin
filliping, we use the coin operator to make superposition on the
coin space, and instead of the walking in CRW we use the
translation operator to move quantum particle according to the
coin's degrees of freedom.

In one-dimensional QW we have two degrees of freedom in the coin
space $H_c$, spanned by $\{|L\rangle,|R\rangle\}$, and infinite
degrees of freedom in the position space $H_p$, spanned by
$\{|i\rangle \,\,\,  i=-\infty,\cdots,\infty \}$. The whole
Hilbert space of the walker is defined as the tensor product of
the coin space $H_c$ and the position space $H_p$, i.e.
$H=H_p\otimes H_c$ . In  one step of the QW we first make
superposition on the coin space with the coin operator $U_c$ and
after that we move the particle according to the coin state with
the translation operator $S$ as follows
\begin{equation}\label{U-w}
U_w=S.(I\otimes U_c)
\end{equation}
where
\begin{equation}\label{S}
S=\sum\limits_x {\left| {x + 1} \right\rangle \left\langle x \right| \otimes \left| R \right\rangle \left\langle R \right| +
 \left| {x - 1} \right\rangle \left\langle x \right| \otimes \left| L \right\rangle \left\langle L
 \right|}.
\end{equation}
Therefore the quantum walking is defined by
\begin{equation}\label{t-step-waking}
|\Psi\left(t+1\right)\rangle={U_w}|\Psi\left(t\right)\rangle \,\,\rightarrow \,\,|\Psi\left(t\right)\rangle={U_w}^t|\Psi\left(0\right)\rangle.
\end{equation}

In ideal case that one can isolate the system from the environment
perfectly, the time evolution of the system takes place coherently
and  the state of the walker remains pure after $t$ steps of the
walking. But in the real world where the interaction between the
system and the environment is unavoidable, the purity of the
system is decreased and the state of the walker becomes mixed. In
this case we should consider the evolution of the whole system,
i.e, system+environment, defined on the  Hilbert space
\begin{equation}\label{HE+HP+HC}
H=H_E\otimes H_p\otimes H_c.
\end{equation}
By definition of the Kraus operators, one step of the walking can
be written as follows
\begin{equation}\label{first-step-rho}
\rho\left(t+1\right)=\sum\limits_{n = 0}^m {E_n \rho\left(t\right)
E_n^\dagger}.
\end{equation}
For $t$ steps,  we can write
\begin{equation}\label{KR-rho-rep-t}
\rho\left(t\right)=\sum\limits_{n_t = 0}^m {...\sum\limits_{n_2 = 0}^m {\sum\limits_{n_1 = 0}^m {E_{n_t}...E_{n_2}E_{n_1}
\rho\left(0\right) E_{n_1}^\dagger E_{n_2}^\dagger ...E_{n_t}^\dagger}}}.
\end{equation}
It is worth to note that the Eq. (\ref{KR-rho-rep-t}) is general
and the Kraus operators $E_n$ include the whole information of all
types of evolution. It follows therefore that the coin operator,
the translation operator and the environment effects all are
embedded in $E_n$ and we didn't assume any restriction yet.

The Kraus operators satisfy an important constraint known as
\textit{completeness relation} \cite{Nielsen} which arise from the
fact that the trace of $\rho\left(t+1\right)$ must be equal to
one, i.e.
\begin{eqnarray}
   1=\Tr\left(\rho\left(t+1\right)\right)&=&\Tr\left(\sum\limits_n{E_n \rho(t) {E_n}^{\dag}}\right)\\ \nonumber
    &=&\Tr\left(\sum\limits_n{{E_n}^{\dag}E_n \rho(t)}\right)
\end{eqnarray}
Since this is true for all $\rho$ then
\begin{equation}\label{compeletness-relation}
   \sum\limits_n{{E_n}^\dag E_n }=I.
\end{equation}
To make any progress we should therefore find the Kruse operators
for our system defined in Eq. (\ref{KR-rho-rep-1}),  and use the
Eq. (\ref{KR-rho-rep-t}) in order to obtain the final state
$\rho(t)$. Evidently,  the $E_n$  are operators that act on the
system (coin+position) Hilbert space, and therefore we can write
the general form of $E_n$ as follows
\begin{eqnarray}\label{general-En}
E_n  &=& \sum\limits_{x,x'}{\sum\limits_{i,j} {a_{x,x',i,j}^{(n)} \left| {x'} \right\rangle \left\langle x \right| \otimes \left| i \right\rangle \left\langle j \right|} }\\\nonumber
&=& \sum\limits_x {\sum\limits_l {\sum\limits_{i,j} {a_{x,l,i,j}^{(n)} \left| {x + l} \right\rangle \left\langle x \right| \otimes \left| i \right\rangle \left\langle j \right|} } }
\end{eqnarray}
where $x,l=-\infty,\cdots,\infty$ and $i,j=\{L,R\}$.\\

%Burn et. al \cite{} has introduced an approach based on Fourier transformation and affine map for investigation coin-decoherence on QW. In this paper we generalize their approach and show that this generalized formalism is applicable for all kinds of decoherence.
In the following section we show that a reasonable suggestion on
the coefficient $a^{(n)}$ of Eq. (\ref{general-En}) enables us  to
derive useful analytical expression for the first and second
moments of position.

\subsection{Environment effects (decoherence)}
In this section we briefly discuss about the interpretation of
noisy evolution and the Kraus representation.  Assume that the
$p_i$ is the probability that the $i$th unknown reason affects the
state of the system. In the other word, the  evolution operators
$A_i$ act on the system with the corresponding probabilities
$p_i$.

Now we are interested in the walker, though the unitary evolution
takes place in the Hilbert space of the system+environment, given
in Eq. (\ref{HE+HP+HC}). Furthermore the environment could be the
rest of the universe, so determining the exact reasons  needs the
investigation of the whole universe which is impossible.

Fortunately, in practice, understanding of the exact form of the
environment is not required. Let us assume that we have $r$
different operators $A_i,\; i=1,\cdots,r$ where each of them acts
on the system with the probability $p_i$. If the environment be in
the state $\ket{e_i}$ then the operator $A_i$ acts on the system.
Therefore we can imagine the $r$-dimensional Hilbert space for the
environment, spanned  by $\{|e_i\rangle\}_{i=1}^{r}$, and the
following initial state for the environment
\begin{equation}\label{env_0}
\left| env_0 \right\rangle  = \sqrt{p_1} \left| e_1 \right\rangle  + \sqrt {p_2} \left| e_2  \right\rangle +...+\sqrt {p_r} \left| e_r  \right\rangle.
\end{equation}
It is clear that the probability of finding the environment in the
state $\ket{e_i}$ is $p_i$.
%the environment will be in the state $\ket{e_i}$ with probability $p_i$.
Therefore we can write the unitary transformation of the whole
system (environment+system) as follows
\begin{equation}\label{general-U}
U = \left| {e_1 } \right\rangle \left\langle {e_1 } \right| \otimes A_1  + \left| {e_2 } \right\rangle \left\langle {e_2 } \right| \otimes A_2  + ... + \left| {e_r } \right\rangle \left\langle {e_r } \right| \otimes A_r.
\end{equation}
So the Kraus operators will be
\begin{equation}\label{E-respectto-A}
E_i  = \left\langle {e_i } \right|U\left| {env_0 } \right\rangle  = \sqrt {p_i} A_i.
\end{equation}
Accordingly, the Eq. (\ref{first-step-rho}) gives the density
matrix after the first step as
\begin{eqnarray}\label{first.step-A}
 \rho ' = \sum\limits_{i = 1}^{r} {E_i \rho E_i^\dag  }=\sum\limits_{i = 1}^{r} {p_i A_i \rho A_i^\dag.  }
\end{eqnarray}
As we see, this is exactly our expectation from the influence of
noise because it gives $\rho '$ as a mixture of different
evolutions with the corresponding probabilities $p_i$. With Eqs.
(\ref{env_0}) and (\ref{general-U}) we can represent the effects
of noise by the Kraus operators.\\

\section{Formalism}
\label{sec:formalism} The Fourier transformation is a powerful
tool for analytical investigation of the one-dimensional QW, where
is defined as follows
\begin{equation} \label{Four-x}
\left| x \right\rangle  = \int\limits_{ - \pi }^\pi
{\frac{{dk}}{{2\pi }}e^{ - ikx} \left| k \right\rangle }.
\end{equation}
Using this transformation we can write  Eq. (\ref{general-En}) in
$k$ space as \small
\begin{equation}\label{En-Kspace}
\tilde{E_n}  = \sum\limits_{x,l} {\sum\limits_{i,j}
{a_{x,l,i,j}^{(n)} \iint {\frac{{dkdk'}}{{4\pi ^2 }}e^{ - ilk} e^{
- ix\left( {k - k'} \right)} \left| k \right\rangle \left\langle
{k'} \right| \otimes \left| i \right\rangle \left\langle j
\right|} } }.
\end{equation}
\normalsize In the following we assume that the coefficients
$a^{(n)}_{x,l,i,j}$ are not dependent  on the coordinate $x$. This
means that the probability of translation in the position space
depends only on the length $l$ of the translation, not on the
position $x$ where  the translation takes place. In view of this
constraint on the $a^{(n)}$, we are able to derive an analytical
expression for the first and second moments of the position, which
is applicable for a large family of decoherence in the
one-dimensional QW. In this regime, the  Eq. (\ref{En-Kspace})
takes the following form \small
\begin{equation}\label{En-kspace-factor-a}
\tilde{E_n}  = \sum\limits_l {\sum\limits_{i,j} {a_{l,i,j}^{(n)} \iint {\frac{{dkdk'}}{{2\pi ^2 }}e^{ - ilk} \delta\left(k-k'\right) \left| k \right\rangle \left\langle {k'} \right| \otimes \left| i \right\rangle \left\langle j \right|} } }
\end{equation}
\normalsize where we have used  the orthonormalization relation
\begin{equation}\label{completeness-x}
\sum\limits_x {e^{ - ix\left( {k - k'} \right)} }  = 2\pi \delta
\left( {k - k'} \right).
\end{equation}
 By integration on $k'$ and
changing the order of integration and summation we get
\begin{equation}\label{En-kspace-finalform}
\tilde{E_n}  = \int {\frac{{dk}}{{2\pi }}\left| k \right\rangle
\left\langle k \right| \otimes C_n\left(k\right)}
\end{equation}
where
\begin{equation}\label{Cn}
C_n \left( k \right) = \sum\limits_l {\sum\limits_{i,j} {a_{l,i,j}^{(n)} e^{ - ilk} \left| i \right\rangle \left\langle j \right|}. }
\end{equation}
Now by writing  the general form of $\rho_0$ in the $k$ space as
\begin{equation} \rho _0  = \iint {\frac{{dkdk'}}{{4\pi ^2
}}\left| k \right\rangle \left\langle {k'}
 \right| \otimes \left| {\psi _0 } \right\rangle \left\langle {\psi _0 } \right|}
\end{equation}
the first step of walking from Eq. (\ref{first-step-rho}) becomes
\begin{eqnarray}\nonumber
\rho '  &=& \iint {\frac{{dkdk'}}{{4\pi ^2 }}\left| k \right\rangle \left\langle {k'} \right| \otimes \sum\limits_n {C_n\left(k\right) \left| {\psi _0 } \right\rangle \left\langle {\psi _0 } \right|C_n^\dag\left(k'\right)  } }  \\
&=&\iint {\frac{{dkdk'}}{{4\pi ^2 }}\left| k \right\rangle \left\langle {k'} \right| \otimes \mathcal{L}_{k,k'}\left| {\psi _0 } \right\rangle \left\langle {\psi _0 } \right|  }
\end{eqnarray}
where $\mathcal{L}_{k,k'}$ is a superoperator defined by
\begin{equation}\label{L-k,k'}
  \mathcal{L}_{k,k'} \tilde{O}= \sum\limits_n{C_n\left(k\right) {\tilde{O}} C_n^\dag\left(k'\right).}
\end{equation}
Therefore after $t$ steps, the state of the walker and the
probability of finding the walker in  position $x$ are,
respectively
\begin{equation}\label{rho-superop-t-step}
\rho\left(t\right)=\iint {\frac{{dkdk'}}{{4\pi ^2 }}\left| k
\right\rangle \left\langle {k'} \right| \otimes
\mathcal{L}^{t}_{k,k'}\left| {\psi _0 } \right\rangle \left\langle
{\psi _0 } \right|}
\end{equation}
and
\begin{eqnarray}\label{probability(x,t)}
p\left( {x,t} \right) &=& \iint {\frac{{dkdk'}}{{4\pi ^2 }}
\scalar{x}{k}\scalar{k'}{x}\Tr\left( {\mathcal{L}_{kk'}^t \rho _0
} \right)}\\ \nonumber &=& \iint {\frac{{dkdk'}}{{4\pi ^2 }}
e^{-ix\left(k'-k\right)}\Tr\left( {\mathcal{L}_{kk'}^t \rho _0 }
\right)}.
\end{eqnarray}
Note that the completeness relation given in Eq.
(\ref{compeletness-relation}) implies that
\begin{equation}\label{completeness-Cn}
 \sum\limits_n{C^{\dag}_n(k) C_n(k)}=I.
\end{equation}
It follows from this condition on the coin operator that the
superoperators  $\mathcal{L}_{k,k}$ are \textit{trace preserving},
i.e.
\begin{eqnarray}
\Tr\left( {\mathcal{L}_{k,k} \tilde O} \right) &=& \Tr\left(
{\sum\limits_n {C_n \left( k \right)\tilde OC_n^\dag  \left( k
\right)} } \right)\\\nonumber
 &=& \Tr\left( {\sum\limits_n {C_n^\dag  \left( k \right)C_n \left( k \right)\tilde O} } \right) = \Tr\left( {\tilde O} \right)
\end{eqnarray}
and therefore
\begin{equation}\label{trace-preserving}
\Tr\left( \mathcal{L}^{m}_{k,k} \tilde O \right) =  \Tr\left(
{\tilde O} \right)
\end{equation}
which are true for arbitrary operator $\tilde{O}$.

Evidently if we know the operators $C_n\left(k\right)$, the state
and the probability distribution of the system can be obtained by
direct using of Eqs. (\ref{rho-superop-t-step}) and
(\ref{probability(x,t)}). But unfortunately it is these equations
which are difficult to handle, and therefore we turn our attention
on the \textit{moments} of this distribution. The $m$th moment of
the probability distribution $p\left(x,t\right)$ is defined by
\begin{equation}\label{m-moment}
\left\langle {x^m } \right\rangle  = \sum\limits_x {x^m p\left( {x,t} \right)}.
\end{equation}
Inserting  Eq. (\ref{probability(x,t)}) in  Eq. (\ref{m-moment})
we get
\begin{equation}\label{moment-m}
\left\langle {x^m } \right\rangle  = \frac{1}{{4\pi ^2
}}\sum\limits_x {x^m \iint {dkdk'e^{ - ix\left( {k' - k} \right)}
\Tr\left( {\mathcal{L}_{kk'}^t \rho _0 } \right)} }.
\end{equation}
Using the orthonormalization  relation (\ref{completeness-x}), we
get for the  moments  $\langle x \rangle$ and $\langle x^2
\rangle$
\begin{eqnarray}\label{moments-1,2}
\left\langle {x } \right\rangle  &=& \frac{-i}{{2\pi  }}\iint
{dkdk' \frac{d\delta\left(k'-k\right)}{dk} \Tr\left( {\mathcal{L}_{kk'}^t \rho _0 }
\right)}  \\ \nonumber \left\langle {x^2 } \right\rangle  &=&
\frac{1}{{2\pi }}\iint {dkdk' \frac{d^2\delta\left(k'-k\right)}{dk'dk} \Tr\left(
{\mathcal{L}_{kk'}^t \rho _0 } \right)}.
\end{eqnarray}
In order to  carry out these integrations we need the following
equations
\begin{eqnarray}\label{diff-L}
\frac{d}{{dk}}\Tr\left( {\mathcal{L}_{kk'} \tilde O} \right) =
\Tr\left(\sum\limits_n { {\frac{{dC_n \left( k
\right)}}{{dk}}\tilde OC_n^\dag  \left( {k'} \right)} }\right)
\\\nonumber \frac{d}{{dk'}}\Tr\left( {\mathcal{L}_{kk'} \tilde O}
\right) = \Tr\left(\sum\limits_n {{C_n\left( {k} \right) \tilde O
\frac{{d C_n^\dag \left( k'\right)}}{{dk'}} } }\right)
\end{eqnarray}
where according to Eq. (\ref{Cn})
\begin{eqnarray}
\frac{{dC_n \left( k \right)}}{{dk}} &=&  - i\sum\limits_l{\sum\limits_{i,j} {la_{l,i,j}^{(n)} e^{ - ilk}
\left| i \right\rangle \left\langle j \right|} }\\\nonumber
\frac{{dC_n^{\dag}\left( k' \right)}}{{dk'}} &=&   i\sum\limits_l{\sum\limits_{i,j} {l\left(a_{l,i,j}^{(n)}\right)^* e^{  ilk'}
 \left| j \right\rangle \left\langle i \right|}}.
\end{eqnarray}
Since the superoperator $\mathcal{L}_{k,k'}$ acts on the density
matrix which  is positive and Hermitian, we can write the Eq.
(\ref{diff-L})  as follows
\begin{eqnarray}\label{diff-L-G-rep}
\frac{d}{{dk}}\Tr\left( {\mathcal{L}_{kk'} \rho} \right) =
\Tr\left( {\mathcal{G}_{kk'} \rho} \right) \\\nonumber
\frac{d}{{dk'}}\Tr\left( {\mathcal{L}_{kk'} \rho} \right)
=\Tr\left( {\mathcal{G^\dag}_{k' k} \rho} \right)
\end{eqnarray}
where
\begin{equation}\label{G}
\mathcal{G}_{k,k'}\tilde{O}=\sum\limits_n {\frac{{dC_n \left( k
\right)}}{{dk}}\tilde{O}C^{\dag}_n \left(k'\right)}.
\end{equation}
Finally by carrying out the integrations of Eq.
(\ref{moments-1,2}) and putting everything together, we arrive at
the following relations for the first and second moments
\begin{widetext}
\begin{eqnarray}\label{moments-1,2-final-form}
{\langle x\rangle} _t &=&  i \int_{ - \pi }^\pi{ \frac{dk}{{2\pi}}
\sum\limits_{m = 1}^t  {\Tr\left\{ {{\cal G}_k \left( {{\cal
L}_k^{m - 1} \left| {\psi _0 \rangle \langle \psi _0 } \right|}
\right)} \right\}} }\\ \nonumber \left\langle {x^2 }
\right\rangle_t  &=& \int_{ - \pi }^\pi
{\frac{dk}{{2\pi}}\sum\limits_{m = 1}^t {\sum\limits_{m' = 1}^{m -
1} {\Tr\left\{ {\mathcal{G}_k^\dag  \mathcal{L}_k^{m - m' - 1}
\left( {\mathcal{G}_k \mathcal{L}_k^{m' - 1}
\left|\psi_0\rangle\langle\psi_0\right| } \right)} + \mathcal{G}_k
\mathcal{L}_k^{m - m' - 1} \left( {\mathcal{G}_k^\dag
\mathcal{L}_k^{m' - 1} \left|\psi_0\rangle\langle\psi_0\right| }
\right)\right\}}}}  \\\nonumber &+& \int_{ - \pi }^\pi
{\frac{dk}{{2\pi}}\sum\limits_{m = 1}^t {\Tr\left\{
{\mathcal{J}_{k}\left( {\mathcal{L}_{k}^{m - 1}
 \left|\psi_0\rangle\langle\psi_0\right| } \right)}  \right\}}}
\end{eqnarray}
\end{widetext}
where we have  defined
\begin{equation}\label{J}
\mathcal{J}_{k}=\left.\frac{{d\mathcal{G}_{k,k'}^\dag
}}{{dk}}\right|_{k'=k} =\left.\sum\limits_n {\frac{{dC_n \left( k
\right)}}{{dk}}\tilde{O}\frac{{dC^{\dag}_n \left( k'
\right)}}{{dk'}}} \right|_{k'=k}
\end{equation}
and  have used $\mathcal{L}_{k}\equiv\mathcal{L}_{k,k}$ and
$\mathcal{G}_{k}\equiv\mathcal{G}_{k,k}$ for the sake of
simplicity. It is worth to note that thus obtained moments in Eq.
(\ref{moments-1,2-final-form}) are a generalization of the moments
given by Brun et.al in \cite{Brun03}, in the sense that they are
applicable for all kinds of decoherence.

%We will find the variance of QW in the presence of broken line noise( coin-position-decoherence) as an example and we will show that in coin-decoherence case these moments reduce to the Brun et.al formula.

\section{Calculation and result}
In this section our task is to show that the method is applicable
for all possible kinds of decoherence. To this aim, in the first
subsection we consider the broken line decoherence as an example
of the coin-position decoherence for which we cannot separate the
coin noise from the position noise. In the second subsection, we
restrict ourselves to the case that the coin is subjected to
decoherence but the position is free of noise, and show that our
method reduces to the result of Brun et al given in \cite{Brun03}.

\subsection{coin-position decohrence}
\begin{figure}[h]
\centering
\includegraphics[width=6cm]{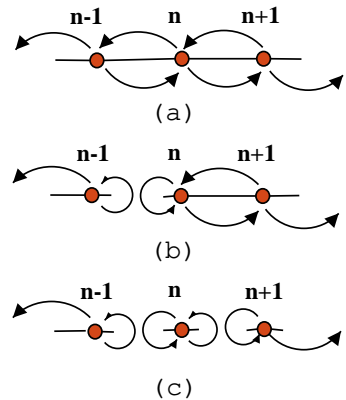}
\caption{(Figure from \cite{deco}) Possible situations for site n
of the line when there are: (a) no broken links, (b) the link to
the left of the site is broken, and (c) both links are broken. The
arrows indicate the direction of the probability flux associated
to the L,R components.} \label{fig:links}
\end{figure}

In this subsection we investigate the effects of coin-position
decoherence on the QW. As an example of our approach we find the
diffusion coefficient of the one-dimensional QW in the presence of
broken line noise. This model has been studied by Romanelli et al
\cite{deco}, where the authors have calculated the diffusion
coefficient numerically. They have introduced broken line
decoherence for the one-dimensional QW in such a way that the
connection between neighbors of current position breaks with
probability $p$, i.e. the walker doesn't walk with probability
$p$. It turns out that with probability $(1-p)^2$ both links are
connected and the walker proceeds normally. On the other hand with
probability $p(1-p)$ (res. $p^2$) one line (res. both lines) is
broken and the walker is returned back (res. is stopped) (see
Fig.\ref{fig:links}).

Let us rename S in Eq. (\ref{S}) to $S_1$ as the translation
operator for the case that we don't have any broken line. We also
define the translation operators  $S_2,S_3,S_4$ for the cases that
the left,  right and both lines  are broken, respectively, i.e.
\begin{equation}\label{S_i}
\begin{array}{l}
 S_1  = \sum\limits_x {\left| {x + 1} \right\rangle \left\langle x \right| \otimes \left| R \right\rangle \left\langle R \right|
  + e^{i\theta _1 } \left| {x - 1} \right\rangle \left\langle x \right| \otimes \left| L \right\rangle \left\langle L \right|}  \\
 S_2  = \sum\limits_x {\left| {x + 1} \right\rangle \left\langle x \right| \otimes \left| R \right\rangle \left\langle R \right|
 + e^{i\theta _2 } \left| x \right\rangle \left\langle x \right| \otimes \left| R \right\rangle \left\langle L \right|}  \\
 S_3  = \sum\limits_x {\left| x \right\rangle \left\langle x \right| \otimes \left| L \right\rangle \left\langle R \right|
 + e^{i\theta _3 } \left| {x - 1} \right\rangle \left\langle x \right| \otimes \left| L \right\rangle \left\langle L \right|}  \\
 S_4  = \sum\limits_x {\left| x \right\rangle \left\langle x \right| \otimes \left( {\left| R \right\rangle \left\langle L \right|
 + e^{i\theta _4 } \left| L \right\rangle \left\langle R \right|} \right)}.
 \end{array}
\end{equation}
The situations corresponding to the above  translation operators
are demonstrated in Fig.\ref{fig:links}. Note that we have used
the general phase for each translation operator. Later we will see
that   we need these generic phases in order to have the correct
physical phenomena.

As mentioned before,  the environment state determines the broken
line situation. For instance if the environment is in the state
$|e_1\rangle$, then $S_1$ acts on the system and so on. Therefore,
according to the Eq. (\ref{env_0}) we can define  the initial
state of the environment as follows
\begin{equation}
\left| {env_0 } \right\rangle  = \left( {1 - p} \right)\left| {e_1
} \right\rangle  + \sqrt {p\left( {1 - p} \right)} \left( {\left|
{e_2 } \right\rangle  + \left| {e_3 } \right\rangle } \right) +
p\left| {e_4 } \right\rangle.
\end{equation}
Also from Eqs. (\ref{U-w}) and (\ref{general-U}), the general
unitary transformation of the system-environment is as follows
\begin{eqnarray}\label{U}
U = \left| {e_1 } \right\rangle \left\langle {e_1} \right| \otimes S_1 \left( {I \otimes H } \right)
+ \left| {e_2 } \right\rangle \left\langle {e_2 } \right| \otimes S_2 \left( {I \otimes H } \right)\\ \nonumber
 + \left| {e_3 } \right\rangle \left\langle {e_3 } \right| \otimes S_3 \left( {I \otimes H } \right)
 + \left| {e_4 } \right\rangle \left\langle {e_4 } \right| \otimes S_4 \left( {I \otimes H } \right)
\end{eqnarray}
where we have assumed the Hadamard walk, i.e. $U_c=H$. Therefore
according to Eq. (\ref{general-U}) $A_i=S_i\left(I\otimes
H\right)$, and we find from Eq. (\ref{E-respectto-A}) the Kraus
operators as
\begin{eqnarray}
\nonumber
E_1  &=& \left( {1 - p} \right)\sum\limits_x  \left| {x + 1} \right\rangle \left\langle x \right| \otimes \left| R \right\rangle
\left\langle R \right|H
\\
&&\,\,\,\,\,\,\,\,\,\,\,\,\,\,\,\,\,\,\,\,\,\,\,\,\,\,\, +
e^{i\theta _1 } \left| {x - 1} \right\rangle \left\langle x
\right| \otimes \left| L \right\rangle \left\langle L \right|H
\\ \nonumber
E_2  &=&\sqrt{p\left( {1 - p} \right)}\sum\limits_x \left| {x + 1} \right\rangle \left\langle x \right| \otimes
\left| R \right\rangle \left\langle R \right|H
\\
&&\,\,\,\,\,\,\,\,\,\,\,\,\,\,\,\,\,\,\,\,\,\,\,\,\,\,\,\,\,\,\,\,\,\,\,\,\,
+ e^{i\theta _2 } \left| x \right\rangle \left\langle x \right|
\otimes \left| R \right\rangle \left\langle L \right|H  \\
\nonumber E_3  &=&\sqrt{p\left( {1 - p} \right)}\sum\limits_x
\left| x \right\rangle \left\langle x \right| \otimes \left| L
\right\rangle \left\langle R \right|H
\\
&&\,\,\,\,\,\,\,\,\,\,\,\,\,\,\,\,\,\,\,\,\,\,\,\,\,\,\,\,\,\,\,\,\,\,
+ e^{i\theta _3 } \left| {x - 1} \right\rangle \left\langle x \right| \otimes \left| L \right\rangle \left\langle L \right|H
\\
E_4  &=&p\sum\limits_x {\left| x \right\rangle \left\langle x \right| \otimes \left( {\left| R \right\rangle \left\langle L \right|H
+ e^{i\theta _4 } \left| L \right\rangle \left\langle R \right|H} \right)}.
\end{eqnarray}
All of these operators are in the form given by Eq.
(\ref{general-En}). For instance,  the coefficients $a^{(1)}$ for
$E_1$ are
\begin{eqnarray}\label{an}
\begin{array}{ll}
a_{1,R,R}^{(1)}  =\frac{{\left( {1 - p} \right)}}{{\sqrt 2 }} \,\,\,\,\,\,\,\,\,\,\,\,\,\,\,& a_{-1,L,L}^{(1)}  =-e^{i\theta_1}\frac{{\left( {1 - p}
\right)}}{{\sqrt 2 }}\\
a_{1,R,L}^{(1)}  = \frac{{\left( {1 - p} \right)}}{{\sqrt 2 }} & a_{-1,L,R}^{(1)}  =e^{i\theta_1} \frac{{\left( {1 - p} \right)}}{{\sqrt 2 }}
\end{array}
\end{eqnarray}
where can be used in Eq. (\ref{Cn}) and obtain $C_1$ as
\begin{equation}
C_1=\frac{{\left( {1 - p} \right)}}{{\sqrt 2 }}\left( {\begin{array}{*{20}c}
   {e^{ - ik} } & {e^{ - ik} }  \\
   {e^{i(k + \theta _1 )} } & { - e^{i(k + \theta _1 )} }  \\
\end{array}} \right).
\end{equation}
Similarly, we can find the other $C_i$ as follows
\begin{eqnarray}
C_2&=&\sqrt {\frac{{\left( {1 - p} \right)}}{2}} \left( {\begin{array}{*{20}c}
   {e^{ - ik}  + e^{  i\theta _2 } } & {e^{ - ik}  - e^{  i\theta _2 } }  \\
   0 & 0  \\
\end{array}} \right)\\\nonumber
C_3&=&\sqrt {\frac{{\left( {1 - p} \right)}}{2}} \left( {\begin{array}{*{20}c}
   0 & 0  \\
   {1 + e^{i\left( {k + \theta _3 } \right)} } & {1 - e^{i\left( {k + \theta _3 } \right)} }  \\
\end{array}} \right)\\\nonumber
C_4&=&\frac{p}{{\sqrt 2 }}\left( {\begin{array}{*{20}c}
   1 & { - 1}  \\
   {e^{i\theta _4 } } & {e^{i\theta _4 } }  \\
\end{array}} \right).
\end{eqnarray}
These $C_i$ must satisfy the completeness relation given by  Eq.
(\ref{completeness-Cn}). After some calculations we find that for
any $\theta_1 , \theta_4$ but only when $\theta_2-\theta_3=\pi$,
the Eq. (\ref{completeness-Cn}) is satisfied. In the following,
for simplicity, we choose $\theta_1=\theta_3=\theta_4=0$ and
$\theta_2=\pi$.

According to Eq. (\ref{L-k,k'}), with these $C_i$ in hands, we can
find the superoperator $\mathcal{L}_{k,k'}$.  But as we know, we
have to calculate the $m$th power of  this superoperator which is
not an easy task to handle.  To get around this complexity, we
follow the method of Ref. \cite{Brun03} and use the \textit{affine map}
approach. Since the $\mathcal{L}_k$ is linear we can represent it
as a matrix acting on the space of two-by-two operators. To do
this we note that one can represent any two-by-two matrix by a
four-dimensional column vector as
\begin{equation}\label{rho-affine map}
\tilde{O} = r_0I+r_1\sigma_1+r_2\sigma_2+r_3\sigma_3\equiv\left(
{\begin{array}{*{20}c}
   {r_0 }  \\
   {r_1 }  \\
   {r_2 }  \\
   {r_3 }  \\
\end{array}} \right)
\end{equation}
where
\begin{equation}
r_i=\frac{1}{2}\Tr\left(\sigma_i\tilde{O}\right).
\end{equation}
Here we have defined $\sigma_0=I$, and $\sigma_i$ ($i=1,2,3$) are
the usual Pauli matrices.

Now, in order to find  the superoperators $\mathcal{L}_k ,
\mathcal{G}_k ,\mathcal{G}^{\dag}_k$ and $\mathcal{J}_k$, we
represent the action  of them on an arbitrary two-by-two matrix
$\tilde{O}$ as follows
\begin{equation}\label{L-matrix-rep}
\mathcal{L}_k\tilde{O}\equiv\left( \begin {array}{*{20}c} 1&0&0&0\\
0&0& e & f+p^2\\
0&0& -f+p^2& e \\
0&1-2p& -2g & -2h
\end {array} \right) \left( {\begin{array}{*{20}c}
   {r_0 }  \\
   {r_1 }  \\
   {r_2 }  \\
   {r_3 }  \\
\end{array}} \right)
\end{equation}
\begin{equation}\label{G-matrix-rep}
\mathcal{G}_{k}\tilde{O}\equiv\left( \begin {array}{*{20}c} 0&i \left( p-1 \right) &ig &ih\\
0&0& f & -e \\
0&0& e & f\\
i\left(p-1\right)&0& -h & g \end {array}
 \right) \left( {\begin{array}{*{20}c}
   {r_0 }  \\
   {r_1 }  \\
   {r_2 }  \\
   {r_3 }  \\
\end{array}} \right)
\end{equation}
\begin{equation}\label{J-matrix-rep}
\mathcal{J}_k\tilde{O}\equiv\left( \begin {array}{*{20}c} 1-p&0&0&0\\
0&0& -e & -f\\
0&0& f & -e \\
0&1-p& 0 & 0 \end {array} \right) \left( {\begin{array}{*{20}c}
   {r_0 }  \\
   {r_1 }  \\
   {r_2 }  \\
   {r_3 }  \\
\end{array}} \right)
\end{equation}
and
\begin{equation}\label{G^dag=G^*}
\mathcal{G}_k^{\dag}\tilde{O}=\mathcal{G}_k^{*}\tilde{O}.
\end{equation}
where the last is obtained, simply,  from the Hermiticity of the
Pauli matrices. Here $e,f,g$ and $h$ are functions of $p$ and $k$,
defined by
\begin{eqnarray}\nonumber
e(p,k)&=&(p-1)^2 \sin \left(2k \right)\\\nonumber
f(p,k)&=&\left(p-1 \right)^2  \cos \left( 2k \right)\\\nonumber
g(p,k)&=&p\left(1-p \right) \sin \left( k \right)\\
h(p,k)&=&p\left(1-p \right) \cos \left( k \right).
\end{eqnarray}
With this representation we can calculate the moments given by Eq.
(\ref{moments-1,2-final-form}). Since
$\Tr\left(\mathcal{L}_{k}\tilde{O}\right)=2r_0$ and the
$\mathcal{L}_k$ is trace preserving, so it doesn't change $r_0$.
Therefore the only nontrivial result arise from the following
three-by-three submatrix
\begin{equation}\label{M-matrix-rep}
M_k=\left( \begin {array}{*{20}c}
0& e & f+p^2\\
0& -f+p^2& e \\
1-2p& -2g & -2h\end {array} \right).
\end{equation}
For finding $\langle x \rangle$ in Eq.
(\ref{moments-1,2-final-form}) we should take trace, accordingly,
only the first row of $\mathcal{G}_{k}$ is important
\begin{equation}
\langle x\rangle _t  = \frac{-1}{{\pi }}\int_{ - \pi }^\pi  {dk\left( {\begin{array}{*{20}c}
   {\left( {p - 1} \right)} & g & h  \\
\end{array}} \right)\left[ {\sum\limits_{m = 1}^t {M_k^{m - 1} } } \right]\left( {\begin{array}{*{20}c}
   {r_1 }  \\
   {r_2 }  \\
   {r_3 }  \\
\end{array}} \right)}.
\end{equation}
We can use the geometric progression to simplify the summation in
this equation.
\begin{equation}\label{moments-1-M-rep}
\langle x\rangle _t \approx  \frac{-1}{{\pi }}\int_{ - \pi }^\pi  {dk\left( {\begin{array}{*{20}c}
   {\left( {p - 1} \right)} & {g} & {h}  \\
\end{array}} \right)\left[ \left(I-M_k\right)^{-1}\right]\left( {\begin{array}{*{20}c}
   {r_1 }  \\
   {r_2 }  \\
   {r_3 }  \\
\end{array}} \right)}.
\end{equation}
Note that we omit the $M_k^t$ in this equation because all
eigenvalues of $M_k$ obey $0<\left|\lambda\right|<1$ and $M^t_k
\to 0$ in long time limit. All $t$ dependence has vanished,
therefore in long time limit the first moment is time independent.

The matrix $\left(I-M_k\right)$  is exactly invertible and
calculation of $\langle x \rangle_t$ is straightforward but our
interest is finding the diffusion coefficient with below
definition
\begin{equation}\label{D}
D = \frac{1}{2}\mathop {\lim }\limits_{t \to \infty } \frac{{\partial \sigma ^2 }}{{\partial t}}
\end{equation}
where $\sigma^2=\langle x^2 \rangle-\langle x \rangle^2$. Clearly
the time independent term doesn't contribute to $D$, therefore we
focus on finding the second moment in Eq.
(\ref{moments-1,2-final-form}). To begin with, let us first
calculate the last term of $\langle x^2 \rangle _t$, i.e.
\begin{equation}
 \int_{ - \pi }^\pi {\frac{dk}{{2\pi}}\sum\limits_{m = 1}^t {\Tr\left\{  {\mathcal{J}_{k}\left( {\mathcal{L}_{k}^{m - 1} \left|\psi_0\rangle\langle\psi_0\right| } \right)}  \right\}}}.
\end{equation}
Since only the first row of $\mathcal{J}_{k}$ (Eq.
(\ref{J-matrix-rep})) makes nonzero result in the trace, therefore
\begin{equation}\label{moments-2-last-term}
\int_{ - \pi }^\pi  {\frac{dk}{{\pi}}\sum\limits_{m = 1}^t {\left({\begin{array}{*{20}c}
 {1-p} & { 0 } & {0} &{0} \\
\end{array}}\right) {\mathcal{L}_k^{m' - 1} \left( {\begin{array}{*{20}c}
   {1/2 }  \\
   {r_1 }  \\
   {r_2 }  \\
   {r_3 }  \\
\end{array}} \right)  } }}=\left(1-p\right)t
\end{equation}
where  we put $r_0=1/2$ because of the normalization condition of
$|\psi_0\rangle$. Now we turn our attention to the first term of
$\langle x^2 \rangle_t$ in Eq. (\ref{moments-1,2-final-form} ),
i.e.
\begin{widetext}
\begin{eqnarray}\label{moment-2-first-term}
\nonumber
&=& \int_{ - \pi }^\pi  {\frac{dk}{{2\pi }}\sum\limits_{m = 1}^t {\sum\limits_{m' = 1}^{m - 1} {\Tr\left\{ {\mathcal{G}_k^\dag  \mathcal{L}_k^{m - m' - 1} \left( {\mathcal{G}_k \mathcal{L}_k^{m' - 1} \left|\psi_0\rangle\langle\psi_0\right| } \right)} + \mathcal{G}_k \mathcal{L}_k^{m - m' - 1} \left( {\mathcal{G}_k^\dag  \mathcal{L}_k^{m' - 1} \left|\psi_0\rangle\langle\psi_0\right| } \right)\right\}}}}\\
&=&-i\int_{ - \pi }^\pi  {\frac{dk}{\pi}\sum\limits_{m = 1}^t {\sum\limits_{m' = 1}^{m - 1} \left({\begin{array}{*{20}c}
 {0} & { (p-1) } & {g} & {h}  \\
\end{array}}\right) {\mathcal{L}_k^{m - m' - 1} \left( \mathcal{G}_k -\mathcal{G}_k^{\dag}\right) \mathcal{L}_k^{m' - 1} \left( {\begin{array}{*{20}c}
   {1/2 }  \\
   {r_1 }  \\
   {r_2 }  \\
   {r_3 }  \\
\end{array}} \right)  } }}
\end{eqnarray}
\end{widetext}
where in the second line we use the facts that only the first row
of $\mathcal{G}_k$ and $\mathcal{G}_k^{\dag}$ makes nonzero result
in the  trace and  that the first row of $\mathcal{G}_k$ is pure
imaginary.

Since the $\mathcal{L}_{k}$ is trace preserving and leaves $r_0$
unchanged, we write
\begin{equation}\label{act-L-on-rho0}
\mathcal{L}_k^{m'-1}\left( {\begin{array}{*{20}c}
   {1/2 }  \\
   {r_1 }  \\
   {r_2 }  \\
   {r_3 }  \\
\end{array}} \right) =\left( {\begin{array}{*{20}c}
   {1/2 }  \\
   {r'_1 }  \\
   {r'_2 }  \\
   {r'_3 }  \\
\end{array}} \right).
\end{equation}
Also from Eqs. (\ref{G-matrix-rep}) and (\ref{G^dag=G^*}) we find
the exact form of $\mathcal{G}_{k}-\mathcal{G}^{\dag}_{k}$ as
\begin{equation}\label{G-Gd}
 \left( \mathcal{G}_{k}-\mathcal{G}^{\dag}_{k}\right) \tilde{O} =2i\left( \begin {array}{*{20}c} 0&\left( p-1 \right) & g & h \\
0&0&0&0\\
0&0&0&\\
\left(p-1\right)&0&0&0\end {array}
 \right) \left( {\begin{array}{*{20}c}
   {r_0 }  \\
   {r_1 }  \\
   {r_2 }  \\
   {r_3 }  \\
\end{array}} \right).
\end{equation}
Therefore we can easily write
\begin{equation}
\left( \mathcal{G}_{k}-\mathcal{G}^{\dag}_{k}\right)\mathcal{L}_k^{m'-1}\left( {\begin{array}{*{20}c}
   {1/2 }  \\
   {r_1 }  \\
   {r_2 }  \\
   {r_3 }  \\
\end{array}} \right) =i\left( {\begin{array}{*{20}c}
   {f\left(k,p,\vec{r'}\right) }  \\
   {0 }  \\
   {0 }  \\
   {p-1}  \\
\end{array}} \right).
\end{equation}
Now putting everything together, we get for the second moment
$\langle x^2 \rangle_t$
\begin{eqnarray}\label{moment-2-befor-M-rep}
{\langle x^2 \rangle}_t&=&\left(1-p\right)t+\int_{ - \pi }^\pi  {\frac{dk}{{\pi }} \Gamma\left(k,p,t\right)}
\end{eqnarray}
where (we drop out arguments for simplicity)
\begin{equation}\label{Gamma}
\Gamma=\sum\limits_{m = 1}^t {\sum\limits_{m' = 1}^{m - 1} \left({\begin{array}{*{20}c}
 {0} & { (p-1) } & {g} & {h}  \\
\end{array}}\right) {\mathcal{L}_k^{m - m' - 1} \left( {\begin{array}{*{20}c}
   {0}  \\
   {0 }  \\
   {0 }  \\
   {p-1 }  \\
\end{array}} \right)  } }.
\end{equation}
Note that the $f\left(k,p,\vec{r'}\right)$ doesn't appear in
$\Gamma$, since
\begin{equation}
  \left({\begin{array}{*{20}c}
 {0} & { (p-1) } & {g} & {h}  \\
\end{array}}\right) {\mathcal{L}_k^{m - m' - 1} \left( {\begin{array}{*{20}c}
   {f\left(k,p,\vec{r'}\right)}  \\
   {0 }  \\
   {0 }  \\
   {0 }  \\
\end{array}} \right)  }=0.
\end{equation}

Same as before we can use the matrix $M_k$ for calculating Eq.
(\ref{Gamma})
\begin{equation}\label{Gamma-M-representation}
\Gamma=\left({\begin{array}{*{20}c}
 { (p-1) } & {g} & {h}  \\
\end{array}}\right) \left[\sum\limits_{m = 1}^t {\sum\limits_{m' = 1}^{m - 1} {M_k^{m - m' - 1}}}\right] \left( {\begin{array}{*{20}c}
   {0 }  \\
   {0 }  \\
   {p-1 }  \\
\end{array}} \right)
\end{equation}
where
\begin{eqnarray}\label{moment-2-M-simplification}
\sum\limits_{m = 1}^t {\sum\limits_{m' = 1}^{m - 1} {M_k^{m - m' - 1}}}
&=&\left(I-M_k\right)^{-1} \sum\limits_{m = 1}^t {\left(I-M_k^m\right)}\\\nonumber
&=&\left(I-M_k\right)^{-1}\left\{t-\left(I-M_k\right)^{-1}M\right\}.
\end{eqnarray}
By inserting this into Eq. (\ref{Gamma-M-representation}), the
calculation of $\langle x^2 \rangle_t$ will be straightforward.

\begin{figure}
\centering
\includegraphics[width=7cm]{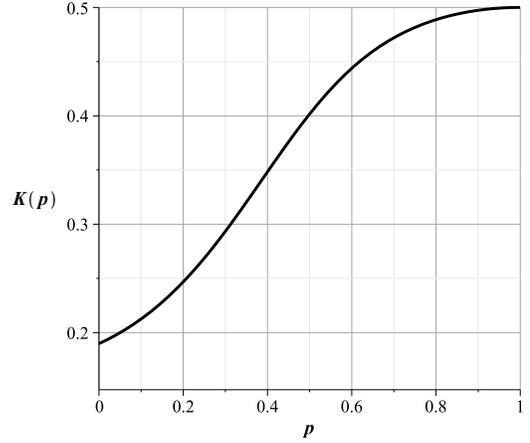}
\caption{The K(p) versus p} \label{fig:k}
\end{figure}

Our interest is finding $D$ here, so only the time dependent terms
will be important for us. By inserting the time dependent term of
Eq. (\ref{moment-2-M-simplification}) into Eq.
(\ref{moment-2-befor-M-rep}) and  definition of $D$ given in Eq.
(\ref{D}), we have
\begin{eqnarray}\label{D-partial-Gamma}
\nonumber
D &=& \frac{1}{2}\mathop {\lim }\limits_{t \to \infty } \frac{{\partial \sigma ^2 }}{{\partial t}}\\
&=&\frac{1}{2}\left\{\left(1-p\right)+\int_{ - \pi }^\pi  {\frac{dk}{\pi} \frac{\partial \Gamma}{\partial t}  }\right\}.
\end{eqnarray}
The matrix $\left(I-M_k\right)^{-1}$ is exactly invertible but
it's elements are a little long to appear here. Fortunately this
matrix appear in the form of
\begin{equation}\label{Gamma-partial}
\frac{\partial\Gamma}{\partial t}=\left({\begin{array}{*{20}c}
 { (p-1) } & {g} & {h}  \\
\end{array}}\right) \left(I-M_k\right)^{-1} \left( {\begin{array}{*{20}c}
   {0 }  \\
   {0 }  \\
   {p-1 }  \\
\end{array}} \right)
\end{equation}
i.e., it is sandwiched  between two vectors. It is not difficult
to calculate  Eq. (\ref{D-partial-Gamma}) to have
\begin{eqnarray}\label{D-final-p}
\nonumber
D &=& \frac{1}{2}\left\{(1-p)+\frac{(1-p)^2}{p} \left[  1-I\left( 1-p\right) \right]\right\}\\
&=& \frac{1-p}{p}K\left(p\right)
\end{eqnarray}
where
\begin{equation}\label{K(p)}
  K\left(p\right)=\frac{1}{2}\left\{1-\left(1-p\right)I(x)\left( 1-p\right)\right\}
\end{equation}
and
\begin{equation}\label{I}
I\left(x\right) = \int_{-\pi}^\pi {\frac{dk}{2\pi} \frac{\cos \left( k \right)+x}{\cos \left( k \right)^2x+\cos
 \left( k \right)x+ 2x^2-2x+1}}.
\end{equation}
The expression of $D\left(p\right)$ in Eq. (\ref{D-final-p}) is
exactly same as the numerical result of Ref. \cite{deco} in which
the authors have  proposed a constant value for $K$ and have
estimated $K \approx 0.4$ by linear regression. But our analytical
calculations show that the coefficient $K$ is a function of $p$
(Eq. (\ref{K(p)})) and ranges from 0.19 to 0.5  as $p$ goes from 0
to 1 (see Fig. \ref{fig:k} ).
The numeric prediction of Romanelli et al implies that for p=0.44
the diffusion coefficient is exactly the same as the unbiased
classical random walk, i.e. 1/2, but our analytical calculations
indicate  that this probability is less than the numeric
prediction and is about p=0.417. This means that for $p<0.417$ the
diffusion coefficient is greater than the classical one and the
quantum walker spreads more faster than the classical one. The
origin of this difference is behind of the fact that the numerical
method can never be applied in problems with infinite steps.

\subsection{coin decoherence}
In this subsection we  show that in the coin decoherence, the
general formalism of Eq. (\ref{moments-1,2-final-form}) reduces to
the formalism of Brun et al presented in \cite{Brun03}.

Let us suppose that  before each step of walking, the operators
$D_c^{(n)}$ acts on the coin subspace with probability $p_n$. Then
according to Eq. (\ref{general-U}) the $A_n$ are
\begin{equation}
A_n=S.\left(I \otimes H \right).\left(I \otimes D_c^{(n)}\right)=S.\left(I \otimes HD_c^{(n)} \right).
\end{equation}
By plugging  it into Eq. (\ref{E-respectto-A}) and using the
explicit form of $S$ we get
\begin{eqnarray}\label{E_n-coin}
\nonumber
E_n=\sqrt{p_n}\sum\limits_x &&\left| {x + 1} \right\rangle \left\langle x \right| \otimes \left| R \right\rangle \left\langle R \right|HD_c^{(n)} \\
&+&  \left| {x - 1} \right\rangle \left\langle x \right| \otimes \left| L \right\rangle \left\langle L \right|HD_c^{(n)}.
\end{eqnarray}
Now by writing the operator $HD_c^{(n)}$ in the basis
$\{\left|R\right\rangle ,\left|L\right\rangle\}$
\begin{equation}
HD_c^{(n)}=\sum\limits_{r,s=\{R,L\}}{\gamma_{r,s}^{(n)}\left|r\right\rangle\left\langle s \right|}
\end{equation}
the Eq. (\ref{E_n-coin}) takes the following form
\begin{eqnarray}\label{E_n-coin-HD-explicit}
\nonumber
E_n=\sqrt{p_n}\sum\limits_x \sum\limits_{s=\{R,L\}}&\gamma_{R,s}^{(n)}&{\left| {x + 1} \right\rangle \left\langle x \right| \otimes \left| R \right\rangle \left\langle s \right|} \\
+&\gamma_{L,s}^{(n)}&  \left| {x - 1} \right\rangle \left\langle x \right| \otimes \left| L \right\rangle \left\langle s \right|.
\end{eqnarray}
This equation is exactly in the form of Eq. (\ref{general-En})
with the specified values for $l$, i.e.  $l=1,-1$, hence
\begin{equation}\label{a-in-coin-only}
a^{(n)}_{1,R,s}=\gamma^{(n)}_{R,s}  \,\,\,\,\,,\,\,\,\,
a^{(n)}_{-1,L,s}=\gamma^{(n)}_{L,s}.
\end{equation}
Therefore according to Eq. (\ref{Cn}) we have
\begin{equation}\label{Cn-coin-only}
C_n \left( k \right) = \sum\limits_{j=\{L,R\}} {\gamma_{R,j}^{(n)} e^{ - ik} \left| R \right\rangle \left\langle j \right|+\gamma_{L,j}^{(n)} e^{  ik} \left| L \right\rangle \left\langle j \right|}
\end{equation}
and
\begin{eqnarray}\label{diffk-Cn-coin-only}
\nonumber
\frac{dC_n \left( k \right)}{dk} &=& -i\sum\limits_{j=\{L,R\}} {\gamma_{R,j}^{(n)} e^{ - ik} \left| R \right\rangle \left\langle j \right|-\gamma_{L,j}^{(n)} e^{  ik} \left| L \right\rangle \left\langle j \right|} \\
&=&-iZC_n \left( k \right).
\end{eqnarray}
And finally according to the definition of $\mathcal{L}_{k,k'}$ in
Eq. (\ref{G})
\begin{equation}\label{G_-iZL}
\mathcal{G}_{k,k'}\tilde{O}=-iZ\left(\mathcal{L}_{k,k'}\tilde{O}\right).
\end{equation}
Clearly
$\mathcal{G}^{\dag}_{k,k'}\tilde{O}=i\left(\mathcal{L}_{k,k'}\tilde{O}\right)Z$,
and the $\mathcal{J}_k$ in Eq. (\ref{J}) will be
\begin{equation}\label{J-coin-only}
\mathcal{J}_{k}\tilde{O}=\left.\frac{{d\mathcal{G}_{k,k'}^\dag \tilde{O}  }}{{dk}}\right|_{k'=k}=Z\left(\mathcal{L}_{k}\tilde{O}\right)Z.
\end{equation}
Inserting Eq. (\ref{J-coin-only}) into the last term of Eq.
(\ref{moments-1,2-final-form}) we get
\begin{eqnarray}\label{J-equal-t-coin-only}
\nonumber &&\int_{ - \pi }^\pi{\frac{dk}{{2\pi }}\sum\limits_{m =
1}^t {\Tr\left\{  {\mathcal{J}_{k}\left( {\mathcal{L}_{k}^{m - 1}
\left|\psi_0\rangle\langle\psi_0\right| } \right)}
\right\}}}\\\nonumber
&=&\int_{ - \pi }^\pi {\frac{dk}{{2\pi }}\sum\limits_{m = 1}^t {\Tr\left\{  Z\left[\mathcal{L}_{k}\left( {\mathcal{L}_{k}^{m - 1} \left|\psi_0\rangle\langle\psi_0\right| } \right)\right]Z\right\}}}\\
&=&\int_{ - \pi }^\pi {\frac{dk}{{2\pi }} \sum\limits_{m = 1}^t
{\Tr\left\{ \mathcal{L}_{k}^{m}
\left|\psi_0\rangle\langle\psi_0\right| \right\}}}=t.
\end{eqnarray}
Finally, putting Eqs. (\ref{G_-iZL}) and
(\ref{J-equal-t-coin-only}) in the moment expressions of Eq.
(\ref{moments-1,2-final-form}), reduces them to the same
expression as introduced by Brun et al \cite{Brun03} for only coin
decoherence, i.e.
\begin{widetext}
\begin{eqnarray}\label{moments-1,2-final-form-coin}
{\langle x\rangle} _t &=&  \int_{ - \pi }^\pi{\sum\limits_{m =
1}^t  {\frac{dk}{{2\pi }} \Tr\left\{ Z\mathcal{L}_k^{m} \left|
{\psi _0 \rangle \langle \psi _0 } \right|\right\}}} \\ \nonumber
\left\langle {x^2 } \right\rangle_t  &=& t+\int_{ - \pi }^\pi
{\frac{dk}{{2\pi }}\sum\limits_{m = 1}^t {\sum\limits_{m' = 1}^{m
- 1} {\Tr\left\{Z  \mathcal{L}_k^{m - m'} \left( {Z
\mathcal{L}_k^{m'} \left|\psi_0\rangle\langle\psi_0\right| }
\right)\right\} + \Tr \left\{ Z \mathcal{L}_k^{m - m'} \left(
{\left(\mathcal{L}_k^{m'}
\left|\psi_0\rangle\langle\psi_0\right|\right)Z }
\right)\right\}}}}.
\end{eqnarray}
\end{widetext}
Therefor the first and second moments given in Eq.
(\ref{moments-1,2-final-form}) are general in the sense that thay
can be applied  for any types of decoherence, including
coin-position decoherence, coin decoherence and position
decoherence.

\section{Summary and Conclusion}
We have present an analytical method for calculating the first and
second moments of the one-dimensional QW, which are applicable for
all kinds of decoherence. Our method is a generalization of the
Brun et al method, introduced  in \cite{Brun03}, where there the
authors  have presented  the analytical expressions for the first
and second moments in the presence of coin decoherence and have
shown that the transition from quantum to classic happens even for
the weak coin noise.

We have made exact calculations on the broken line decoherence as
an example of the non-separable decoherence  of the coin-position
system. Our analytical calculations are in consistent  with the
previous numerical result \cite{deco}. We have shown that the $K$
in the diffusion coefficient $D$ is, indeed,  a function of $p$,
and ranges  from 0.19 to 0.5 as $p$ varies from 0 to 1.

%*************************** add after ********************************
Furthermore our calculation shows that for $p=0.417$ the value of
$D$ is $1/2$, exactly the same as the unbiased classical random
walk. This specific value of $p$ is critical in the sense that the
transition from quantum to classic happens at this point. In other
word, the quantum walk with non zero broken line probability,
spreads faster than the classical random walk if the probability
of broken line is  not higher than 0.417. For higher values of
$p$, the  lines are broken too frequently and this will prevent
the full diffusion. The critical $p$, estimated by the numerical
calculation of Ref. \cite{deco}, is a little higher than the one
obtained here analytically. The difference arises from the
non-ability of numerical method in simulation of the QW with
infinite steps.

%***********************************************************************************
%, and we found
%that for $p\approx 0.42$, the diffusion coefficient is exactly the
%same as the unbiased classical random walk. {\bf It means that for
%smaller broken line frequencies the diffusion rate of the QW is
%higher than the classical rate.}
We have also studied the case that only the coin is subjected to
decoherence  and have shown that the expressions for the first and
second moments lead to the Brun et al results presented in
\cite{Brun03}.

%****************************** add after *******************************************

%{\bf We also investigate effects of position only decoherence on
%probability distribution and variance as a final example of
%possible kinds of decoherence. We refer authors to [OUR TUNNELING
%PAPER] to see complete calculation of position only decoherence

The long time behavior of moments  and, of course, the variance
can be used as a qualitative measure of classicality. Therefore
these calculated moments provides a powerful tool for
investigating the one-dimensional QW in the presence of any kinds
of decoherence. For instance, finding the particular cases for
which dependency of the variance on the time remains quadratic,
i.e. the decoherence does not affect the quantumness, is an
application of these moments. Brun et al \cite{Brun03} have shown
that the quadratic time  dependency of the variance is preserved
for multiple-coin QW subjected to the coin decoherence, but their
calculation is restricted to the coin decoherence only. In this
paper this restriction is relaxed and the model allows the
presence  of any kind of decoherence.

These formulas are useful also in the experimental implementation
of the quantum walk.  The application of our results for two kinds
of decoherence, i.e. the coin-only and the coin-position noises,
have been done in this paper. The application  on the position
decoherence have been also tested but the full calculations were
so long to appear here. The results of our calculations on the
tunnelling effect, for which the position is subjected to
decoherence, and it's consistency  with the previous numerical
work \cite{Dur-prop} confirm the usefulness  of these expressions
for the position decoherence too.  Other aspects of the tunnelling
effects are also under consideration. The paper, therefore, can be
regarded as a further development in the study of  decoherent
one-dimensional QW.

\acknowledgements{M.A. thanks  G. Abal for valuable discussions.}

%********************************************************************************

% {\bf For instance the
%analytical calculations of the tunnelling effect in the QW, in
%which the position is subjected to decoherence, the result of
%[...........] have shown that the position decoherence preserves
%the quadratic time dependency of the variance.}

% Bibtex
% El argumento a \bibliographystyle se refiere a un fichero style.bst, el cual define cómo se mostrarán las citas
% Los estilos estándar distribuidos con BibTeX son:
% alpha : ordenación alfabética. Las etiquetas se forman con el nombre del autor y el año de publicación.
% plain  : ordenación alfabética. Las etiquetas son numéricas.
% unsrt : como plain, pero las entradas por orden de cita.
% abbrv  : como plain, pero con etiquetas más compactas.
\bibliography{qw}

\begin{thebibliography}{10}

\bibitem{Shenvi03}
N.~Shenvi, J.~Kempe, and B.~Whaley,
\newblock Phys. Rev. A {\bf 67}, 52307 (2003), eprint quant-ph/0210064.

\bibitem{Ambainis03}
A.~Ambainis,
\newblock SIAM Journal on Computing {\bf 37}, 210 (2007), preprint
  quant-ph/0311001.

\bibitem{AKR}
A.~Ambainis, J.~Kempe, and A.~Rivosh,
\newblock Proc. 16th ACM-SIAM SODA , p. 1099 (2005), quant-ph/0402107.

\bibitem{Ambainis05}
A.~Ambainis,
\newblock SIGACT News {\bf 35}, 22 (2004), preprint quant-ph/0504012.

\bibitem{Childs04}
A.~M. Childs and J.~Goldstone,
\newblock Phys. Rev. A {\bf 70}, 022314 (2004), preprint quant-ph/0306054.

\bibitem{Tulsi08}
A.~Tulsi,
\newblock Phys. Rev. A {\bf 78}, 012310 (2008), quant-ph/0806.1257.

\bibitem{Childs2009}
A.~M. Childs,
\newblock Phys. Rev. Lett. {\bf 102}, 180501 (2009).

\bibitem{Nayak}
A.~Nayak and A.~Vishwanath,
\newblock preprint quant-ph/0010117.

\bibitem{FG98}
E.~Farhi and S.~Gutmann,
\newblock Phys. Rev. A {\bf 58}, 915 (1998).

\bibitem{Childs08}
A.~M. Childs,
\newblock preprint quant-ph/0810.0312.

\bibitem{Strauch}
F.~Strauch,
\newblock Phys. Rev. A {\bf 74}, 030301 (R) (2006), preprint quant-ph/0606050.

\bibitem{qw-markov}
A.~Romanelli {\em et~al.},
\newblock Phys. A {\bf 338}, 395 (2004), e-print quant-ph/0310171.

\bibitem{Konno}
N.~Konno,
\newblock J. Quant. Inf. Proc. {\bf 1}, 345 (2002), quant-ph/0206053.

\bibitem{Venegas}
S.~Venegas-Andraca, J.~Ball, K.~Burnett, and S.~Bose,
\newblock New J. Phys. {\bf 7}, 221 (2005), preprint quant-ph/0411151.

\bibitem{cycle1}
M.~Bednarska, A.~Grudka, P.~Kurzynski, T.~Luczak, and A.~Wojcik,
\newblock Physics Letters A {\bf 317}, 21 (2003), quant-ph/0304113.

\bibitem{cycle2}
W.~Adamczak {\em et~al.},
\newblock International Journal of Quantum Information {\bf 5}, 781 (2007),
  quant-ph/0708.2096.

\bibitem{Mackay}
T.~T.D.~Mackay, S.~Bartlett, L.~Stephenson, and B.~Sanders,
\newblock Journal of Physics A {\bf 35}, 2745 (2002), preprint
  quant-ph/0108004.

\bibitem{Carneiro}
I.~Carneiro {\em et~al.},
\newblock New J. Phys. {\bf 7}, 156 (2005), preprint quant-ph/0504042.

\bibitem{Omar}
Y.~Omar, N.~Paunkovic, L.~Sheridan, and S.~Bose,
\newblock Phys. Rev. A {\bf 74}, 42304 (2006), preprint quant-ph/0411065.

\bibitem{Amanda}
A.~Oliveira, R.~Portugal, and R.~Donangelo,
\newblock Phys. Rev. A {\bf 74}, 12312 (2006).

\bibitem{Konno2}
K.~Watabe, N.~Kobayashi, M.~Katori, and N.~Konno,
\newblock PRA {\bf 77}, 062331 (2008), quant-ph/0802.2749.

\bibitem{Moore}
C.~Moore and A.~Russell,
\newblock Quantum walks on the hypercube,
\newblock in {\em Proceedings of 6th International Workshop on Randomization
  and Approximation Techniques (RANDOM 2002), Lecture Notes in Computer Science
  2483 (LNCS)}, edited by J.~D.~P. Rolim and S.~Vadhan, pp. 164--178,
  Cambridge, MA, 2002, Springer-Verlag, Berlin, 2002, preprint
  quant-ph/0104137v1.

\bibitem{Marquezino}
F.~Marquezino, R.~Portugal, G.~Abal, and R.~Donangelo,
\newblock PRA {\bf 77}, 042312 (2008), quant-ph/0712.0625.

\bibitem{Chaobin}
C.~Liu and N.~Petulante,
\newblock Phys. Rev. A {\bf 79}, 032312 (2009), quant-ph/0807.2263.

\bibitem{Abal06-ent}
G.~Abal, R.~Siri, A.~Romanelli, and R.~Donangelo,
\newblock Phys. Rev. A {\bf 73}, 042302, 069905(E) (2006), preprint
  quant-ph/0507264.

\bibitem{annabestani}
M.~Annabestani, M.~R. Abolhasani, and G.~Abal,
\newblock (2009), quant-ph/0901.1188.

\bibitem{Goyal}
S.~K. Goyal and C.~M. Chandrashekar,
\newblock quant-ph/0901.0671.

\bibitem{Venegas-entanglementgeneration}
S.~E. Venegas-Andraca and S.~Bose,
\newblock 2009 , quant-ph/0901.3946.

\bibitem{Travaglione}
B.~Travaglione and G.~Milburn,
\newblock Phys. Rev. A {\bf 65}, 032310 (2002).

\bibitem{Knight-OC}
P.~Knight, E.~Roldán, and J.~Sipe,
\newblock Op. Comm. {\bf 227}, 147 (2003).

\bibitem{Dur-prop}
W.~D\"{u}r, V.~Kendon, and H.~Briegel,
\newblock Phys. Rev. A {\bf 66}, 52319 (2002), arXiv preprint quant-ph/0207137.

\bibitem{Du}
J.~Du {\em et~al.},
\newblock Phys. Rev. A {\bf 67}, 042316 (2003), preprint quant-ph/0203120.

\bibitem{Ryan}
C.~Ryan, M.~Laforest, J.~Boileau, and R.~Laflamme,
\newblock Phys. Rev. A {\bf 72}, 062317 (2005), preprint quant-ph/0507267.

\bibitem{Zhao-prop}
Z.~Zhao, J.~Du, H.~Yang, Z.~Chen, and J.~Pan,
\newblock (2002), arXiv preprint quant-ph/0212149.

\bibitem{Kempe03}
J.~Kempe,
\newblock Contemp. Phys. {\bf 44}, 307 (2003), preprint quant-ph/0303081.

\bibitem{KT03}
V.~Kendon and B.~Tregenna,
\newblock Phys. Rev. A {\bf 67}, 42315 (2003), eprint quant-ph/0209005.

\bibitem{Brun03}
T.~Brun, H.~Carteret, and A.~Ambainis,
\newblock Phys. Rev. A {\bf 67}, 32304 (2003), arXiv preprint quant-ph/0210180.

\bibitem{Brun2}
T.~Brun, H.~Carteret, and A.~Ambainis,
\newblock Phys. Rev. Lett. {\bf 91}, 130602 (2003).

\bibitem{lopez}
C.~C. L\'opez and J.~P. Paz,
\newblock Phys. Rev. A {\bf 68}, 052305 (2003).

\bibitem{deco}
A.~Romanelli, R.~Siri, G.~Abal, A.~Auyuanet, and R.~Donangelo,
\newblock Phys. A {\bf 347}, 137 (2004), preprint quant-ph/0403192.

\bibitem{Kendon2003}
V.~Kendon and B.~Tregenna,
\newblock Decoherence in discrete quantum walks,
\newblock in {\em Decoherence and Entropy in Complex Systems}, , Lecture Notes
  in Physics Vol. 633/2003, pp. 253--267, Springer Berlin / Heidelberg, 2003.

\bibitem{Nielsen}
M.~Nielsen and I.~Chuang,
\newblock {\em Quantum Computation and Quantum Information} (Cambridge
  University Press, Cambridge, 2000).

\end{thebibliography}
\bibliographystyle{h-physrev} % phys rev style with eprint

\end{document}